Player Skill Decomposition in Multiplayer Online Battle Arenas

Zhengxing Chen[1][*], Yizhou Sun[2][**], Magy Seif El-Nasr[3][*], and Truong-Huy D. Nguyen[4][***]

[*]Northeastern University

[**]University of California, Los Angeles

[***]Texas A&M University-Commerce

Affiliation

Abstract

Successful analysis of player skills in video games has important impacts on the process of enhancing player experience without undermining their continuous skill development. Moreover, player skill analysis becomes more intriguing in team-based video games because such form of study can help discover useful factors in effective team formation. In this paper, we consider the problem of skill decomposition in MOBA (MultiPlayer Online Battle Arena) games, with the goal to understand what player skill factors are essential for the outcome of a game match. To understand the construct of MOBA player skills, we utilize various skill-based predictive models to decompose player skills into interpretative parts, the impact of which are assessed in statistical terms. We apply this analysis approach on two widely known MOBAs, namely League of Legends (LoL) and Defense of the Ancients 2 (DOTA2). The finding is that base skills of in-game avatars, base skills of players, and players' champion-specific skills are three prominent skill components influencing LoL's match outcomes, while those of DOTA2 are mainly impacted by in-game avatars' base skills but not much by the other two.

Player Skill Decomposition in Multiplayer Online Battle Arenas

# Introduction

Recently a unique type of sports, namely electronic sports (*eSports*), emerges as a popular genre of computer games, in which human players compete with one another in online, simulated environments governed by rules and regulations similar to those found in traditional forms of sports. eSports is a rapidly growing video game market attracting tremendous amounts of professional players, fanatical audiences and tournament organizers. A recent report released by SuperData (2016) showed that the worldwide market for eSports, by the end of 2015, has reached approximately 748 million dollars and is expected to grow to 1.9 billion dollars by 2019.

One of the most played game genres in eSports, Multiplayer Online Battle Arenas or *MOBAs* feature 5-vs-5 competitive matches where each of the ten players selects a champion[5] (or in-game avatar) to combat before each match starts. Each team consisting of five players has a base to defend and the goal is to attack the opposite teams' champions and ultimately destroy the opponent's base. MOBA games are typically characterized by rich choices of champions and dynamic strategies for both individual champion build-up and team cooperation, offering endless replayability to players and introducing sophisticated strategic decision making. MOBA is a highly skill-based game genre, in which good performance and high ranks can only be achieved by players with consistently superior gaming capabilities. Therefore, understanding how players train and build up their skill sets for competitive gaming is extremely important on many fronts.

First, it helps game companies to determine and evaluate different design choices to ensure a fair and engaging environments for players, one of the fundamental aspects of game design, as suggested by Adams (2013). Understanding how a new game element affects players' current skills and performance is crucial in the process of adding variability and updates to a game that is growing old. Second, better understanding of player skill constructs allows more appropriate player matching, which is currently an important game mode that most MOBAs offer to players to enhance play experience, as pointed out by Nguyen, Chen, and El-Nasr (2015). Salen and Zimmerman (2004) found out that matching players with too wide skill gaps leads to completely dominant wins that bore the winners and discourage the losers. Lastly, such knowledge enables game producers to identify and prevent cases of cheats and hacks, which negatively impact the play experience of other legitimate gamers. Duh and Chen (2009) reported that the use of malicious software support has become an inevitable and recurring issue in online gaming. Understanding how player skills are formed and sharpened over time may pave the way towards more effective identification of genuine from fake skills. Within the scope of this paper, we address the problem of unpacking the essential components comprising MOBA games' play skills. Taking as input a data set that records game outcome statistics, we first utilize different statistical models to capture skill decomposition schemes. Each model represents a hypothesis and breakdown of player skill structures, the effectiveness of which directly depends on how accurate it can predict the game outcome. By assessing the difference in terms of predictive power of the hypotheses, the contributions of skill components to players' performance in the game are quantified and validated. This approach does not aim to generate a one-size-fits-all skill model; the goal is to inform users of possible components that make up player skills in a MOBA game and, in statistical terms, the interplay of these components in a player's performance.

---

[5]In League of Legends, in-game avatars selected by players are termed "champion". Other comparable MOBA games may phrase in-game avatars in different words.

The paper makes its contribution in the following aspects:

1. it provides a new context for skill analysis in the MOBA game setting, where player skills are constituted from multi-fold aspects;

2. it details a novel model-based skill analysis approach based on multiple performance forecasters; and

3. it demonstrates the utility of the approach in real-life data sets, which leads to interesting findings on in two of the most popular MOBA games, i.e., League of Legends (LoL) and Defense of the Ancients 2 (DOTA2).

In the rest of the paper, we first present research related to our work and some preliminary pertinent to the MOBA game context and our analysis approach. Next, the proposed skill analysis methodology is laid out in details, starting with data collection, skill structure hypothesis construction, and ending with hypothesis modeling and statistical comparison. The last sections demonstrate the utility of our proposed methodology using the case study of LoL and DOTA2, followed by numerical results and discussions, before the paper is concluded with directions for future work.

## Previous Works

**Skill Rating and Modeling**

The study of skill rating aims to rank players after observing outcomes of their matchups. Bradley-Terry model proposed by Bradley and Terry (1952) was developed to deal with repeated pairwise comparisons among a group of subjects. In Bradley-Terry model, each player is assumed to have a fixed skill scalar and the winning probability of a player is proportional to his skill in the sum of skills of both players involved. The Elo (1978) system, a probabilistic model designed for chess player skill ratings, was developed later in which player skills are assumed to be a random variable following one dimension Gaussian distribution. In the Elo system, player skill means get updated depending on the extent of agreement between expected outcomes and real outcomes. For example, a low skill player beating a high skill player yields a large update in adjusting their skill means closer. However, the variance of player skills is assumed to be a fixed constant. Glicko system designed by Glickman (1999), a Bayesian ranking rating system, was later introduced to model the belief about a player's skill. As he plays increasingly number of games, the belief about his skill becomes stronger. None of Bradley-Terry model, the Elo system or Glicko system was initially applicable to team-oriented games until works from Huang, Lin, and Weng (2004), Menke and Martinez (2008) and Herbrich, Minka, and Graepel (2006) generalize these models. For example, TrueSkill system designed by Herbrich et al. (2006) extends the Elo system to games with flexible sizes and numbers of teams. While these advanced skill rating systems give reasonable player rankings in an efficient manner, they use relatively simple representation, usually single scalars, for player skills and are not tailored specifically for MOBA games. Skill modeling belongs to player modeling, which have been studied for long time by researchers such as Yannakakis, Spronck, Loiacono, and André (2013), Bakkes, Spronck, and van Lankveld (2012) and Charles et al. (2005). In player modeling, characteristics of human players such as strategies, preferences and skills are detected, modeled, predicted and expressed. Skill modeling differs with skill rating in that it aims to encompass player skills in multiple facets and does not necessarily need to rank players. Stanescu (2011) and Chen and Joachims (2016) modelled player skills in multi-dimensions (such as offensive and defensive abilities) but their works are currently only applicable to 1-vs-1 games rather than team based games like MOBA. There are skill models proposed for specific game genres, such as Delalleau et al. (2012) for FPS games and Avontuur, Spronck, and Van Zaanen (2013) for RTS games. To our best knowledge, however, we have seen few skill models in MOBA games. Skill rating/modeling are also intriguing topics in crowdsourcing where a large body of research focus on finding most suitable individuals to complete micro-tasks to maximize outcomes. Among those researches only Kittur (2010), Roy, Lykourentzou, Thirumuruganathan, Amer-Yahia, and Das (2015) and Rahman, Thirumuruganathan, Roy, Amer-Yahia, and Das (2015) consider team based tasks. Our context is still different because our "task" involves two teams, instead of one, in a competitive environment and more factors besides player skills can affect match outcomes.

**MOBA Game Research and Outcome Prediction**

Our paper replies on performance of match outcome prediction to verify proposed skill components. Match outcome prediction is one important area investigating the fundamental elements that determine results of matches. Mislak and Deja Myślak and Deja (2014) tried to predict LoL outcomes by identifying in each team how many players play champions in their personally favored lanes. However, the model was only evaluated on partial matches where team compositions agree with standard norms. Moreover, the performance of the model drops if the two teams have similar numbers of players in their most familiar lanes. Pobiedina et al. Pobiedina, Neidhardt, Calatrava Moreno, Grad-Gyenge, and Werthner (2013) and Pobiedina, Neidhardt, Calatrava Moreno, and Werthner (2013) discovered that four factors contribute to team success namely team composition of champions, the number of friends, the player experience and the conformity of player nationalities. While their focus is not on player skill analysis, their first finding, the team composition of champions, reconciles with our finding that the base skills of champions is one prominent part of player skill formation.

Several works predict match outcomes by harnessing in-game information. In order to predict victory side, Yang, Brent, and Roberts (2014) retrieved characteristics of in-game combats; Mahlmann, Schubert, and Drachen (2016) utilized information from encounters (i.e., situations "when two or more champions from opposing teams are in the range of affect each other"); Drachen et al. (2014) and Rioult, Metivier, Helleu, Scelles, and Durand (2014) exploited spatio information of champions. However, the kind of information is not consistently available in the two games we are investigating (which are the most two popular MOBA games suggested by Minotti (2016)) and our work is from a higher level to examine skill decomposition. Therefore our model does not use in-game information currently.

## PRELIMINARY

We first introduce common rules of MOBA games especially those related to our context. Next, we describe the data forms that our proposed framework is designed to process. Finally, we show parameters and notations necessary for illustrating our model and hypotheses.

**MOBA Game Rules**

In this paper we focus on a largely played MOBA game mode called 5-vs-5 ranked matches of solo queue. A typical such match involves ten players who form two teams of five. Each player selects in turn a champion to combat before the real match starts. Champions, each designed with unique attributes, are expected to have different play styles and tactics in matches. For example, there are champions called fighters who have short-ranged attack ability and excel at surviving combats. Another type of champions are called supports who are weak when alone but can revive allies and slow down opponent movement. Each team has a base to defend and the common victory goal is to destroy the opponents' base. Generally, players adopt strategies not only based on the selected champions but also on the champions selected by teammates and the opponents. It is worth noting that each match is independent of each other in the sense that any accumulated golds, equipment or abilities of champions in one match will not be transferred to any future match, i.e., every player starts each match from scratch.

**Parameters and Notations**

Given the data set that contains such information as described in the previous section, suppose we have $Z$ matches, in which there appeared $M$ champions and $N$ players in total. We use $C_1, C_2, \cdots, C_M$ to index the champions and use $P_1, P_2, \cdots, P_N$ to index the players. We use $\mathcal{M} = \{M_1, M_2, \cdots, M_Z\}$ to denote the outcomes of the $Z$ matches. The two teams in a match are differentiated by red and blue colors. $M_z = 1$ if the red team won over the blue team in match $z$ otherwise $M_z = 0$. The indicator function $I(z, t, C_i, P_j) = 1$ iff $P_j$ selected $C_i$ in match $z$'s team $t$, where $t$ can take values either $r$ or $b$ denoting the red or the blue team respectively. Otherwise, $I(z, t, C_i, P_j) = 0$. Note that, according to the rules of the ranked games we collected, $C_i$ cannot be chosen by two different players in a match $z$. Based on different hypotheses introduced in next section, we construct feature vector $D_z \in \mathbb{R}^K$ for each match $z$. We use $D_{z,i}$ to denote the i-th component of the vector $D_z$. We use $\mathcal{D} = \{D_1, D_2, \cdots, D_Z\}$ to denote the feature vectors for all $Z$ matches.

## Analysis Methodology

In this section, we describe the steps in our analysis methodology, including (1) data crawling and cleaning, in order to acquire usable information of matches; (2) hypothesis construction to form conjectures about player skill formation; (3) training of predictive models constructed based on different hypotheses resulted from Step 2; and (4) hypothesis evaluation and comparison to draw conclusions of skill composition.

**Step 1: Data Crawling and Cleaning**

Using commonly available match statistics data, our approach is to correlate different models and hypotheses of players' skills and their match outcomes. Specifically, we are interested in the game outcome data recorded from 5 vs 5 MOBA matches played by a set of players in a certain range of time, with a set of champions appearing as selected. The matches should be of similar competitive settings and contain every match played by the player set in the period of time. In more details, each match consists of the following information: (1) IDs of the players involved; (2) IDs of the champions involved (3) Win/loss match outcome. The data is readily available for most MOBA games, to the public or not, including LoL and DOTA2. After the data crawling phase, we index matches by integers in $[1, Z]$ and transform player IDs and champion IDs into continuous indices in $[1, N]$ and $[1, M]$ respectively.

**Step 2: Hypothesis Construction**

Next, various skill decomposition hypotheses are utilized to reflect feasible conjectures about player skill formation. We conducted a survey among 8 MOBA players to extract helpful information in hypothesis construction. The survey included the following questions:

1. What is your experience in your played MOBA game?

2. Please list factors that you think could have influence on match outcomes in MOBA games.

3. Please elaborate on the factors you list above which are related to player skills.

The interviewees include 5 veteran players: 3 players are from top 4 tiers in LoL and 2 players with considerably high scores ($\geq 3500$) in solo queue MMR (one measurement of skill) in DOTA2. Among their answers, we extracted three common influential factors in match outcomes that are associated with player skills:

1. Players' general knowledge and experience in controlling match paces and cooperating in team combats.

2. Averagely speaking, champions have different curves to be mastered and when champions confronts in lanes 1-on-1, certain champions have advantages over the others.

3. Even veteran players can master only a couple of champions. Therefore champion-specific skills should be accounted.

In the end, we propose 3 skill components in the skill models: player base skill, champion base skill, and champion-player specific skill. We construct 7 hypotheses to verify them.

Hypothesis 1 ($H_1$): The skill of a player is a single underlying weight denoting his base skill level. In this hypothesis, player skills are regardless of selected champions.

$H_2$: Player skill is not related to individual players but only related to selected champions. Each champion is associated with a single weight, which we call champion base skill, denoting the average influence of which the champion affects match outcomes. Champion base skill can also be interpreted as the average skill of players when selecting the same champions. A player's skill is solely the champion base skill of his selected champion. The team skill is the sum of the weights of the 5 champions selected.

$H_3$: Player skill varies both across players and selected champions. Therefore, we need to estimate champion-specific skill for every player. The team skill is the sum of the 5 player-champion specific skills.

We also explore combinatorial interaction among the three types of skill. As above, if without mentioned specifically, all the hypotheses use the sum of individual skill compositions to represent team skill level.

$H_4$: Player skill consists of player base skill and champion base skill.

$H_5$: Player skill consists of player base skill and his champion specific skill.

$H_6$: Player skill consists of champion base skill and his champion specific skill.

$H_7$: Player skill consists of player base skill, champion base skill and his champion specific skill.

**Step 3: Predictive Models and Training**

We incorporate the 7 hypotheses into multiple predictive models that predicts the game outcomes. The idea is that if a hypothesis is correct, the models incorporating corresponding skill components should demonstrate improved accuracy compared to those without incorporating those skill components.

**Logistic Regression.** We selected logistic regression (LR) to model the hypotheses for the following reasons: (1) LR is a commonly used model for binary classification problem with decent computation time; (2) in preliminaries exploration, we also explored using other common supervised learning models such as Random Forest proposed byLiaw and Wiener (2002), Naive Bayes and SVM proposed by Cortes and Vapnik (1995). However, the first two did not consistently outperform LR and SVM took too long to finish our predictive task; (3) as discussed in the following subsection, the weights (model parameters to be trained) of logistic regression models can be associated with individual players' skills. As such, logistic regression models team performance as the linear summation of individual skills, a typical and effective approach used by Rahman et al. (2015), Delalleau et al. (2012) and Herbrich et al. (2006) to model team level skills based on individuals.

Logistic Regression models the conditional distribution of a match outcome given the match feature vector as a Bernoulli distribution, i.e.,

$$M_z|D_z; \theta \sim Bernoulli(\phi) \quad (1)$$

$$\phi = p(M_z = 1|D_z; \theta) = \sigma(\theta^T D_z) \quad (2)$$

where $\theta \in \mathcal{R}^K$ is the model parameter of the same dimension as input $D_z$ and $\sigma(x) = \frac{1}{1+exp(-\theta^T x)}$ is a sigmoid function.

Logistic regression learns model parameter $\theta \in \mathbb{R}^K$ that maximizes the log likelihood function:

$$\begin{aligned} \ell(\theta) &= \log p(\mathcal{M}|\mathcal{D}; \theta) \\ &= \log \prod_{z=1}^{Z} p(M_z|D_z; \theta) \\ &= \sum_{z=1}^{Z} M_z \log p(M_z|D_z; \theta) + \\ &\quad \sum_{z=1}^{Z} (1 - M_z) \log(1 - p(M_z|D_z; \theta)) \end{aligned} \quad (3)$$

In practice, the objective function of logistic regression is often $\ell(\theta)$ added with a $L_2$ regularization term $C \cdot \theta^T \theta$ to prevent from the overfitting problem, where $C$ is a configurable constant controlling the strength of regularization. For each predictive model, we will use cross validation in order to determine best value of $C$ which helps the model achieve best accuracy on test datasets.

**Feature Vector Construction.**  We need to construct $D_z$, the feature vector of each match $z$, to feed as input to the predictive model. Each of the following paragraphs, starting with the name of its associated predictive model, describes the way to construct $D_z$ to incorporate the 7 proposed hypotheses. We assume $D_z$ in all models as a sparse vector, i.e., components unless mentioned are 0.

**LR-P**[6]: $D_z \in \mathcal{R}^N$. $D_{z,j} = 1$ if and only if $I(z, r, C_i, P_j) = 1$ and $D_{z,j} = -1$ if and only if $I(z, b, C_i, P_j) = 1$. In this way, each component of $\theta$ will correspond to a player's skill weight and $\theta^T D_z$ is exactly the difference of the player base skill sums between the two teams.

**LR-C**: In a similar fashion as in *LR-Player*, there are five 1's and five -1's in $D_z$ denoting the five champions selected by the players on the red team and the blue team respectively.

**LR-P-C** $D_z \in \mathcal{R}^{N+M}$ where the first $N$ components are constructed as *LR-P* and the last $M$ components are constructed as *LR-C*.

**LR-PC**: $D_z \in \mathbb{R}^{N \times M}$ (every player has $M$ weights denoting his skills in the $M$ champions). For $i, j$ such that $I(z, r, C_i, P_j) = 1$, set $D_{z, j \cdot M + i} = 1$. Similarly, for $i, j$ such that $I(z, b, C_i, P_j) = 1$, set $D_{z, j \cdot M + i} = -1$.

**LR-P-C-PC** $D_z \in \mathbb{R}^{N+M+N \times M}$ as the horizontal concatenation of the constructed $D_z$ from *LR-P*, *LR-C* and *LR-PC*.

**Step 4: Hypothesis Comparison**

In order to verify the 7 hypotheses, it is sufficient to use and compare the results of *LR-P*, *LR-C*, *LR-P-C*, *LR-P-C-PC* and a naive baseline (e.g. majority class prediction). For example, $H_3$, which says player skills only consist of champion specific skills, can be verified if: (1) *LR-P-C-PC* improves prediction accuracy significantly over *LR-P-C*; and (2) neither *LR-P* or *LR-C* outperforms the naive baseline. More examples of relying comparisons to verify hypotheses are detailed in *Results and Discussions* section.

We utilize model comparison metrics to reliably demonstrate the superiority of one model over the others with statistical significance. Specifically, when we compare two models with test accuracy means and standard deviations $(mean1, std1)$ and $(mean2, std2)$ respectively, we consider the first one has significant improvement over the second if $mean1 - mean2 > 2(std1 + std2)$[7]. A type of skill composition will be identified as prominent if adding it to feature vectors consistently help predictive models outperform those without incorporating it.

---

[6]LR-P stands for (L)ogistic (R)egression with (P)layer feature components. In the remaining hypotheses, we also use C and PC to represent champion base skills and players' champion specific skills respectively.

## Data Collection

In this paper, we are interested in analyzing player skills in two most popular MOBA games worldwide according to Minotti (2016), League of Legends (LoL) and Defense of Ancients 2 (DOTA2).

In both games, we only consider ranked matches from top tier players as the input data. Ranked matches are those automatically balanced by the matchmaking systems so as to form two teams with similar winning probability prior to game start. Ranked matches are usually played seriously because players can earn recognizable in-game points and titles through winning ranked matches. For this paper we focus on investigating ranked matches because it is the only game mode that provides public available data API in LoL.

Moreover, top tier (most experienced and skilled) players are selected for two reasons: (1) their skills are considered more stable than rookie players therefore it can reduce noise for our predictive model. (2) As we know, in the real world, the population of players usually decreases exponentially as the skill levels goes up. By selecting top tier players we control the number of weights in a computationally acceptable range on single machines given that our focus on skill decomposition rather than large-scale skill rating.

**League of Legends**

As Tassi (2016) and Minotti (2016) reported, LoL is the most played MOBA game in the world, with 90 million summoner names registered, 27 million unique daily players, 7.5 million concurrent users, and 67 million monthly players. In LoL, players can be rewarded (or punished) by winning (or losing) ranked matches in terms of *League Point*. League Points are directly related to their public honored titles (termed *tier*) that a large number of players pursue. The higher League Point a player has, the more esteemed tier he is crowned. League Points as well as tiers are refreshed at the beginning of every year when a new game season starts.

---

[7]Alternatively, one can use paired t-test to check the significance of improvement between the means of the two test accuracy populations. We found in our study both methods lead to the same conclusions in all pairs of model comparisons.

We collected basic information of 972 players (e.g., player ID, registered name, etc) of the top two tiers (CHALLENGER and MASTER) on North America Server from a player database website[8] in September 2015 when most players have played more than half year to establish their tiers since the 2015 season. We used the public official APIs[9] to access detailed data (e.g., win/loss and selected champions) of 231212 unique ranked matches played by the seed players in 2015. We finalized the player set of 93,098 players by including all players appearing in the crawled matches. On the other hand, there are 129 champions played in the matches collected.

**Defense of the Ancient 2 (DOTA2)**

DOTA2 is another popular MOBA game developed by Valve Corporation. In DOTA, the score to reflect player experience and skill is called Match Making Rating (MMR). DOTA2 differs with LoL in that player data including MMR is by default not public available unless the player configures his profile to be public visible. Therefore, in a collected DOTA2 match there might be several players with one identity - anonymous player. In the training phase, we treat all anonymous players as one universal anonymous player. The weights associated with the universal anonymous player are considered as an average representation for all anonymous players.

We downloaded 3.5 million DOTA2 matches played in 2015 from a DOTA2 data statistic website [10]. We only kept ranked matches in ALL PICK, RANDOM DRAFT or CAPTAIN game modes played on North America regions. The three game modes are more similar than others to the competitive setting in the LoL dataset. To filter the matches played by top players, we further selected matches with at least one player with public visible solo-queue MMR larger than 4000 [11]. Ultimately, there are 117,874 players, 1 universal anonymous

---

[8]http://www.lolking.net/leaderboards#/na/1
[9]https://developer.riotgames.com/api/methods
[10]https://yasp.co/blog/35
[11]Players with more than 4000 MMR are approximately top 10% skilled players according to an online curated

player and 111 champions appearing in the 195,538 ranked matches.

## Results and Discussions

The final results are shown in Table 1. We used grid search plus 10-fold cross validation to determine the best regularization strength, $C$, of each logistic regression model. $C$ in the grid search was sampled from log uniform distribution between 0.001 and 1000. After properly tuning of $C$, we report each model with statistics of test accuracy population (mean and standard deviation) obtained in the 10-fold cross validation. We also report the results of two baselines (*TrueSkill* and *BL-MC*). In *TrueSkill*, we sort all matches in an ascending order of match created time. In the first 90% matches, we update player skill ratings according to TrueSkill algorithm proposed by Herbrich et al. (2006), a popular skill rating model. Then in the hold-out 10% latest matches, we fixed the learned player skill ratings to make outcome prediction. In *BL-MC*, we predict match outcomes by the majority labels in each dataset.

**Does the base skills of champions matter?** In both games, the test accuracy of *LR-C* is significantly higher than *BL-MC*. We also observe that *LR-P-C* has noticable improvement over *LR-P*. Therefore we confirm that the base skills of champions is one source of player skill composition.

**Does the base skills of players matter? Does the champion specific skills of players matter?** The two questions can be analyzed together because the results show disagreement between the two games in both questions.

In LoL, the two types of skills are indispensable components in player skill composition. The base skills of players are important because: (1) *LR-P* alone shows non-trivial 3.51% higher test accuracy than *BL-MC*. (2) *LR-P-C* has significant improvement over *LR-C* indicating that the added features, the base skills of players, are non-negligible besides the base skills of champions. (3) *TrueSkill*, which purely models player base skills, has similar accuracy as *LR-P*. Players' champion-specific skills are also important in LoL because *LR-P-C-PC* achieves the best accuracy among all models.

---

global player MMR list: https://github.com/yasp-dota/yasp/wiki/MMR-Data

In DOTA2, the story is much different. First, the base skills of players do not show explicit existence because (1) *LR-P* has the test accuracy close to *BL-MC*, and (2) *LR-P-C* shows no significant improvement over *LR-C*. Second, players' champion specific skills do not constitute a good component of skill composition either. *LR-P-C-PC* has an non-significant 0.13% boost over test accuracy with *LR-P-C*. Combining with the fact that *LR-P* is merely close to the baseline, we conclude that in DOTA2 only the base skills of champions is the main source of player skill composition.

**Discussion**

We find that in LoL, player skill can be decomposed into three parts, namely the base skill of players, the base skill of champions, and the champion-specific skills of players. On the other hand, player skill in DOTA2 is composed mainly of the base skill of champions. Next, we would like to make comments on the larger implication suggested by these results.

**To game designers and developers.** The divergence of player skill composition between LoL and DOTA2 can be an entrance leading game designers and developers to think further about their design and implementation. We suspect the following reasons that potentially account for the divergence of player skill composition: (1) the match making system in DOTA2 has a better quality than that in LoL, such that the base skills of players are hard to be further differentiated from the ranked matches meant to be fair, balancing and competitive. However, the suspicion fails to explain why champion specific skills are not obvious in DOTA2 since such kind skills are not accountable by the current match making systems using one dimension skill scores. (2) The game design of champion characteristics may differ across the two games. It is worth noting that the base skills of champions has a larger effect to determine outcomes in DOTA2 since *LR-C* achieves more than 4% test accuracy in DOTA2 than in LoL and the gap is far beyond what the standard deviations of the two models can explain. This indicates that champions have a larger impact to game outcomes based upon DOTA2's game design. (3) The anonymity issue in the DOTA2 dataset fails the estimation of player side skills due to the loss of information.

To show that the failure is *not* due to the influence of the anonymous players, we randomly marked the same ratio of the LoL players as in DOTA2 to be anonymous players. (Thus all these selected players are marked as one anonymous player identity.) We re-applied all our models and found that prediction improvement is still significant in the modified LoL dataset when adding the base skills and champion specific skills of players into the models. However, as we expected, the means of test accuracy decrease due to the loss of information introduced by the anonymity. Due to the limited space we do not list the results on this dataset.

**To players** The results shown in this paper can be a useful reference for players regarding skill building and match prognosis. For LoL players, they can learn their skill composition of the three kinds of skills and know which part is their skill advantage/disadvantage. Players can even make rough prognosis about the match right after the ten players have finished selecting champions. This can help players identify, for example, the threats of the opponents according to the order of the base skills of champions. Meanwhile there are more interesting questions we want to answer for players in the future. For example, can simultaneous choices of those champions with high base skills necessarily lead to higher win rate? If two teams have the same sum of player base skill weights, will the player with the highest skill weight determine the outcome? A number of intriguing questions can be derived from this paper and we aim to answer them in future works.

**To human resource managers and crowd source task initiators.** The results also allude important lessons for human resource and crowd sourcing communities, where effective team formation is a hot topic. In such scenario, people want to find people of different expertise to achieve a team-wise goal. It is analogous to hypothesize that team performance relies on individual team members skills forming from the base skills of tasks, the base skills of team members, and their task specific expertise. We are looking for future study to verify the hypothesis.

## Conclusion and Future Works

The paper presents our preliminary findings in analyzing player skill composition in MOBA games. Adopting a model-based analysis approach, we find that player skills in MOBA games could include base skill of player, base skill of champion and player's champion specific skill as three prominent components. In addition, we find LoL has more diverse skill compositions than DOTA2. The result can potentially spark the reflection on game design and implementation, raise awareness of skill strength and weakness among players and allude lessons for human resource management and crowd sourcing.

In the future, we would like to explore player skill on team collaboration. For example, what is a player's performance when teaming with players or champions of certain play styles? We also plan to investigate and compare player skill composition on other video game genres.

PLAYER SKILL DECOMPOSITION IN MULTIPLAYER ONLINE BATTLE ARENAS    20experiments. *Applied Statistics*, 377–394.

Herbrich, R., Minka, T., & Graepel, T. (2006). Trueskill™: A bayesian skill rating system. In (pp. 569–576). Advances in Neural Information Processing Systems.

Huang, T.-K., Lin, C.-J., & Weng, R. C. (2004). A generalized Bradley-Terry model: From group competition to individual skill. In *Advances in neural information processing systems* (pp. 601–608).

Kittur, A. (2010). Crowdsourcing, collaboration and creativity. *ACM Crossroads*, *17*(2), 22–26.

Liaw, A., & Wiener, M. (2002). Classification and regression by randomforest. *R news*, *2*(3), 18–22.

Mahlmann, T., Schubert, M., & Drachen, A. (2016). Esports Analytics Through Encounter Detection. In *Mit sloan sports analytics conference*.

Menke, J. E., & Martinez, T. R. (2008). A Bradley–Terry artificial neural network model for individual ratings in group competitions. *Neural computing and Applications*, *17*(2), 175–186.

Minotti, M. (2016). *Comparing MOBAs: League of Legends vs. Dota 2 vs. Smite vs. Heroes of the Storm.* http://venturebeat.com/2015/07/15/comparing-mobas-league-of-legends-vs-dota-2-vs-smite-vs-heroes-of-the-storm/. (Online; accessed 5-May-2016)

Myślak, M., & Deja, D. (2014). Developing game-structure sensitive matchmaking system for massive-multiplayer online games. In *Social informatics* (pp. 200–208). Springer.

Nguyen, T.-H. D., Chen, Z., & El-Nasr, M. S. (2015). *Analytics-based AI Techniques for Better Gaming Experience* (Vol. 2; S. Rabin, Ed.). Boca Raton, Florida: CRC Press.

Pobiedina, N., Neidhardt, J., Calatrava Moreno, M. d. C., Grad-Gyenge, L., & Werthner, H. (2013). On Successful Team Formation: Statistical Analysis of a Multiplayer Online Game. In *2013 ieee 15th conference on business informatics* (pp. 55–62). IEEE.

Pobiedina, N., Neidhardt, J., Calatrava Moreno, M. d. C., & Werthner, H. (2013). Ranking factors of team success. In *Proceedings of the 22nd international conference on world wide web companion* (pp. 1185–1194).

PLAYER SKILL DECOMPOSITION IN MULTIPLAYER ONLINE BATTLE ARENAS    21

Table 1

*Best Average Test Accuracy by Each Model*

| Model | Test Acc ± std | |
|---|---|---|
| | LoL | DOTA2 |
| LR-P | 56.75% ± 0.24% | 52.62% ± 0.47% |
| LR-C | 55.62% ± 0.36% | 59.16% ± 0.41% |
| LR-P-C | 58.82% ± 0.20% | 59.53% ± 0.39% |
| LR-P-C-PC | 60.24% ± 0.16% | 59.66% ± 0.42% |
| TrueSkill | 55.24% ± 0.16% | 52.05% ± 0.12% |
| BL-MC | 53.22% ± 0.16% | 52.65% ± 0.12% |